# Drag Forces, Neutral Wind and Electric Conductivity Changes in the Ionospheric E Region

Petko Nenovski

University Center for Space Research and Technologies (UCSRT), Sofia University, Sofia, Bulgaria

**Abstract.** The neutrals in the Earth environment are in fact free and subjected to drag forces (by ions). In this study we show that drag or friction forces in the ionosphere−thermosphere system initiate changes in the plasma flow, neutral wind, and the conductivity, as well. Ions and electrons embedded in neutral wind field of velocity $\boldsymbol{u}_0$ acquire drifts perpendicular both to the initial neutral wind velocity and to the ambient magnetic field producing a perpendicular electric current. This perpendicular electric current is defined by a conductivity derived previously and the polarization electric field $\boldsymbol{u}_0 \times \boldsymbol{B}$. Self-consistently, the free neutrals acquires an additional neutral velocity component perpendicular to the initial neutral wind velocity $\boldsymbol{u}_0$. The Pedersen and Hall currents wane within a specific time inversely proportional to neutral-ion collision frequency. These findings are relevant to a better understanding of electric current generation, distribution and closure in weakly ionized plasmas where charged particles (plasma) and neutrals are not bound (free) as the ionosphere and thermosphere in the Earth environment, the solar atmosphere-chromosphere region and even dusty plasmas. The theoretical results supposedly address to recent observations of meso-scale neutral wind rotation in the vicinity of auroral arcs (Kosch et al, 2010).

## 1. Introduction

It is well established that thermospheric neutral circulation is driven by a couple of forces as solar-heating pressure, ion drag, advection (momentum), viscous and Coriolis forces. Ion drag and pressure gradients are normally the dominant driving forces in the ionospheric F region as demonstrated by *Killeen and Roble* (1984).

A principal consequence of the ion−neutral interaction process is the contribution of the neutral wind to the total ionospheric Joule heating via e.g. the polarization field $\boldsymbol{E} = \boldsymbol{u}_0 \times \boldsymbol{B}$, where $\boldsymbol{u}_0$ is the neutral wind velocity. It has been found that ignoring the neutral wind the ionospheric electric field has been underestimated and resulted in an error of up to 60% in the Joule heating calculation, at least in the F layer (*Cierpka et al*, 2000). Observations of the F region and neutral wind have evidenced that the neutral wind dynamo was on average 50% of the magnetospheric electric field and contributes an average 41% of in situ Joule heating (*Aruliah et al.* 2004). The maximum in ionospheric Joule heating however is expected to lie at E region heights and it appears very importantly to quantitatively understand the neutral wind contribution to Joule heating at the E region heights, as well. Recently *Kosch et al* (2010) have investigated, the high-latitude E region neutral wind (meso-scale) interaction with auroral arcs during geomagnetically quiet conditions. The authors have succeeded to demonstrate that auroral precipitation in the form of arcs and associated electrical processes *can* have a dramatic effects on the E region neutral wind field on a spatial meso-scale: an auroral arc causes the wind direction to rotate 90° to flow parallel to the arc within 7−16 min in a region less than ~50 km from the arc. Moreover, after having checked whether this effect is due to changing altitude of the optical emission or to the solar heating pressure and Coriolis forces normally dominant in the E region, *Kosch et al* (2010) concluded that the increased plasma density *and* electric field in the immediate vicinity of the auroral arc and the associated ion drag forces are possible sources of the observed neutral wind rotation. Hence, the ion drag time scale or the equivalent e-folding time in the E layer measured for the first time (*Kosch et al., 2010*) turns out to be within 7–16 min.

The possible consequences of such short time scales of ion−neutral interaction might be crucial. In particular, the Pedersen and Hall conductivities dos not render an account of such drag forces and are, in principle, applicable under a simple condition of immobile neutrals. The neutrals in the Earth environment are, in fact, free and subjected to drag forces (by ions). It is expected that depending on ion−neutral interaction time scales the Pedersen and Hall conductivities are improperly applied to electric currents in the Earth's ionosphere and thermosphere and also, to Joule heating evaluations there.

In this study an analysis of the drag forces due to neutral wind, drag conductivity and their relevance to the Earth's ionosphere and thermosphere is performed. A drag conductivity perpendicular to the neutral wind field of velocity $u_0$ and the static magnetic field $B$ is subsequently introduced and derived. The perpendicular electric current expressed through the drag forces is evaluated quantitatively.

Further, our attempt is to indicate that the ion drag force due to an increased plasma density structure in the ionospheric E region as auroral arcs immersed in uniform neutral flow is a principal source of 90° neutral wind rotation under geomagnetically quiet conditions. Hence, the observed 90° neutral wind rotation can entirely be explained by the ion drag mechanism even without involving a localized electric field produced by conductivity inhomogeneities. Along with a current will emerge to flow perpendicular to the initial neutral velocity $u_0$ and the magnetic field $B$.

## 2. Analysis

Supposedly, the neutrals motion of velocity $u_0$ tends to entrain (through collisions) the charged particles and subsequently this entrainment process finally will evolve into a steady state when all electrons and ions adjust their motion to neutrals motion. The charged particles − ions and electrons, (immersed in the more dense neutral atmosphere) continuously collide with neutrals, effectively exchanging impulse momentum. This inference follows from the impulse momentum exchange process through the drag forces proportional to $m_\alpha \nu_{\alpha n}(u_\alpha - u_0)$, where $m_{\alpha n}$, $\nu_{\alpha n}$ are the mass and collision frequency of species α (α = i, ions and α = e, electrons) (Giraud and Petit, 1978; Ballan, 2008). The relative average velocities $u_i - u_0$ and $u_e - u_0$ decrease monotonically until the respective ion and electron velocities, $u_i$ and $u_e$, adjust the neutral velocity $u_0$. If there were no magnetic field $B$ the average ion and electron velocities $u_i$ and $u_e$ would become equal to $u_0$. Then the electric current along $u_0$ would be cancelled. Because both the plasma and the neutrals are immersed in geomagnetic field $B$, this magnetic field enforces the particle trajectories toward the direction of $u_0 \times B$. Hence, a steady state condition of balanced velocities of $u_i$, $u_e$ and neutrals of velocity $u_n$ would emerge after some characteristic time. Indeed, the initial equations of ions, electrons and neutrals read

$$(1) \quad m_i \frac{d\boldsymbol{v}_i}{dt} = e(\boldsymbol{v}_i \times \boldsymbol{B}) - \tilde{m}_{in}\nu_{in}(\boldsymbol{v}_i - \boldsymbol{u}_n - \boldsymbol{u}_0) - \tilde{m}_{ie}\nu_{ie}(\boldsymbol{v}_i - \boldsymbol{v}_e)$$

$$(2) \quad m_e \frac{d\boldsymbol{v}_e}{dt} = -e(\boldsymbol{v}_e \times \boldsymbol{B}) - \tilde{m}_{en}\nu_{en}(\boldsymbol{v}_e - \boldsymbol{u}_n - \boldsymbol{u}_0) - \tilde{m}_{ei}\nu_{ei}(\boldsymbol{v}_e - \boldsymbol{v}_i)$$

$$(3) \quad m_n \frac{d\boldsymbol{u}_n}{dt} = -\tilde{m}_{ni}\nu_{ni}(\boldsymbol{u}_n - \boldsymbol{v}_i) - \tilde{m}_{ne}\nu_{ne}(\boldsymbol{u}_n - \boldsymbol{v}_e)$$

where $\tilde{m}_{in} = m_i m_n/(m_i + m_n)$, $\tilde{m}_{en} = m_e m_n/(m_e + m_n)$ and $\tilde{m}_{ei} \equiv \tilde{m}_{ie} = m_e m_i/(m_e + m_i)$ so on where $m_i = M$, $m_e = m$, and $m_n$ are the mass of ions, electron and neutrals, respectively; $\nu_{in}$, $\nu_{en}$ are collision frequencies of charged particles of species i and e with neutrals; the collision frequency between electron and ions is $\nu_{ie}$ equal to $\nu_{ei}$. The latter are negligible in comparison with collision with neutral and further neglected. Electric fields are excluded. The initial velocity $\boldsymbol{u}_0$ is static one and due to non-electric forces (as pressure). The magnetic field $\boldsymbol{B}$, plasma and the neutral density and velocity $\boldsymbol{u}_0$ are homogeneous. Hence, only polarization field $\boldsymbol{u} \times \boldsymbol{B}$ due to neutral wind is assumed. It is assumed that the ionospheric plasma is cold, hence, terms of spatial coordinates are neglected and not involved in Eqs. (1-3). For convenience, the neutral velocity is represented as a sum of initial velocity $\boldsymbol{u}_0$ and $\boldsymbol{u}_n$. Drag forces (Giraud and Petit, 1978; Bellan, 2008) are thus introduced by terms $\tilde{m}_{in}\nu_{in}(\boldsymbol{v}_i - \boldsymbol{u}_0 - \boldsymbol{u}_n)$ and $\tilde{m}_{en}\nu_{en}(\boldsymbol{v}_e - \boldsymbol{u}_0 - \boldsymbol{u}_n)$. Further, neutral wind entrains the ions and electrons, i.e. the ion and electron velocities $\boldsymbol{v}_i$ and $\boldsymbol{v}_e$ should depend on $\boldsymbol{u}_0$ and of course, would be split into a component $\boldsymbol{u}_0$ and another one due to the Lorentz and drag forces. An analysis of Eqs. (1-3) will show subsequently that the ion and electron velocities $\boldsymbol{v}_i$ and $\boldsymbol{v}_e$ becomes perpendicular to $\boldsymbol{u}_0$. Therefore, the drag forces in direction of initial velocity $\boldsymbol{u}_0$ reduce to $\tilde{m}_{in}\nu_{in}\boldsymbol{u}_0$ and $\tilde{m}_{en}\nu_{en}\boldsymbol{u}_0$. They are further denoted by $\boldsymbol{F}_i$ and $\boldsymbol{F}_e$, respectively. Under static conditions, after lengthy calculations one obtains

$$(1a) \quad m_i \frac{d\tilde{\boldsymbol{v}}_i}{dt} = e(\boldsymbol{v}_i \times \boldsymbol{B}) + \tilde{m}_{in}\nu_{in}\boldsymbol{u}_0 - \tilde{m}_{eff}\nu_{eff}(\boldsymbol{v}_i - \boldsymbol{v}_e)$$

$$(2a) \quad m_e \frac{d\tilde{\boldsymbol{v}}_e}{dt} = -e(\boldsymbol{v}_e \times \boldsymbol{B}) + \tilde{m}_{en}\nu_{en}\boldsymbol{u}_0 - \tilde{m}_{eff}\nu_{eff}(\boldsymbol{v}_e - \boldsymbol{v}_i),$$

where modified ion and electron velocities $\tilde{\boldsymbol{v}}_i$ and $\tilde{\boldsymbol{v}}_e$ are introduced

$$\tilde{\boldsymbol{v}}_i \equiv \boldsymbol{v}_i - \frac{\tilde{m}_{in}\nu_{in}}{\tilde{m}_{ni}\nu_{ni} + \tilde{m}_{ne}\nu_{ne}}\boldsymbol{u}_n \quad \text{and} \quad \tilde{\boldsymbol{v}}_e \equiv \boldsymbol{v}_e - \frac{\tilde{m}_{en}\nu_{en}}{\tilde{m}_{ni}\nu_{ni} + \tilde{m}_{ne}\nu_{ne}}\boldsymbol{u}_n,$$

the effective collision frequency $\nu_{eff}$ reads:

$$(4) \quad \nu_{eff} = \frac{\nu_{in}\nu_{ne}}{\nu_{ni}m_i/(m_i + m_n) + \nu_{ne}m_e/(m_e + m_n)};$$

and the electron-ion collision frequency is neglected (weakly-ionized plasma conditions) and the effective mass, $m_{eff}$, is defined by:

$$(5) \quad \tilde{m}_{eff} = \frac{m_i m_e m_n}{(m_i + m_n)(m_e + m_n)}.$$

Under interested static conditions the time derivatives in Eqs. (1a) and (2a) are zeroed ($\frac{d}{dt} \to 0$). Introducing a perpendicular current, $\boldsymbol{j}_\perp$ (along the direction of the polarization field $\boldsymbol{u}_0 \times \boldsymbol{B}$) and Hall current, $\boldsymbol{j}_H$ (along the direction $\boldsymbol{u}_0$) and summing (1a) and (2a) one obtains: i) $\boldsymbol{j}_H = 0$ and ii) $\boldsymbol{j}_\perp \times \boldsymbol{B} = -N(\tilde{m}_{in}\nu_{in} + \tilde{m}_{en}\nu_{en})\boldsymbol{u}_0$, where $N$ is the plasma density. This suggests that under static approximation, i) the electric (Hall) current along velocity $\boldsymbol{u}_0$ becomes cancelled, and ii) the Ampere force $\mathbf{j}_\perp \times \mathbf{B}$ due to the perpendicular current $\mathbf{j}_\perp$ becomes balanced by the drag force: $\mathbf{F}_{drag} = \mathbf{F}_i + \mathbf{F}_e = N(\tilde{m}_{in}\nu_{in} + \tilde{m}_{en}\nu_{en})\mathbf{u}_0$.

By rearranging we finally obtain:

$$(6) \qquad \mathbf{j}_\perp = N(\tilde{m}_{in}\nu_{in} + \tilde{m}_{en}\nu_{en})\mathbf{u}_0 \times \frac{\mathbf{B}}{B^2} = \sigma_\perp (\mathbf{u}_0 \times \mathbf{B}).$$

where the drag conductivity $\sigma_\perp$ is:

$$(7) \qquad \sigma_\perp = \frac{Ne}{B}\left(\frac{\nu_{in}}{\Omega_i(1+m_i/m_n)} + \frac{\nu_{en}}{\Omega_e(1+m_e/m_n)}\right)$$

with $\Omega_e = eB/m_e$ and $\Omega_i = eB/m_i$ representing the corresponding cyclotron frequencies and $e$ is the elementary electric charge. This conductivity expression exactly coincides with the perpendicular conductivity expression obtained in (Nenovski, 2014).

### 3. Drag conductivity, neutral wind change and related time scale

The sense of the above expression can be understand by the following consideration: Terms proportional to $N(\tilde{m}_{in}\nu_{in} + \tilde{m}_{en}\nu_{en})\mathbf{u}_0$ here stand for the drag force $\mathbf{F}_{drag}$ ($\mathbf{F}_{drag} = \mathbf{F}_i + \mathbf{F}_e$) that balances the competing Ampere force $\mathbf{j}_\perp \times \mathbf{B}$. Note the drag forces are different for ions and electrons: $\mathbf{F}_i = \tilde{m}_{in}\nu_{in}\mathbf{u}_0$ and $\mathbf{F}_e = \tilde{m}_{en}\nu_{en}\mathbf{u}_0$. In the presence of (static) magnetic field $\mathbf{B}$ the drag forces $\mathbf{F}_{i,e}$ produce drift of electrons and ions perpendicular to $\mathbf{F}_{i,e}$ and $\mathbf{B}$ and, hence, as a result, an electric current will flow perpendicular to $\mathbf{F}_\alpha$ and $\mathbf{B}$. Electric currents due to charged particle drifts perpendicular to the ambient static magnetic fields, are termed as drift electric currents. The known forces producing such particle drifts and associated currents are, for example, magnetic field gradient force $\sim \nabla B$; force due to particle motion along the curved magnetic field line proportional to $(\mathbf{B}\cdot\nabla)\mathbf{B}$; polarization current due to time variable electric field $\mathbf{E}$, etc. (Shkarofsky et al, 1966). All these forces generate electric currents $\mathbf{j}_\perp$ perpendicular to the magnetic field $\mathbf{B}$. All such currents can be incorporated by the following formula:

$$(8) \qquad \mathbf{j}_\perp = \frac{\mathbf{F}\times\mathbf{B}}{qB^2}$$

where force $\mathbf{F}$ stands for the above mentioned forces inherent for plasma immersed in static inhomogeneous magnetic field $\mathbf{B}$ and $q$ – the charge of the particle species, ions or electrons. In particular, when an electric field $\mathbf{E}$ is presumed the full drift current becomes zero, because the drift velocity of ions and electrons is the same. If a motion of neutrals of velocity $\mathbf{u}_0$ is assumed, a drag force $\mathbf{F}_{i,e}$ emerges due to collisions exerted on charged particles by neutrals. According to (8) drag forces $\mathbf{F}_i$ and $\mathbf{F}_i$ will produce drift current, as well. Let us evaluate it. Replacing $\mathbf{F}_{i,e}$ equal to $\tilde{m}_{in}\nu_{in}\mathbf{u}_0$ and $\tilde{m}_{en}\nu_{en}\mathbf{u}_0$ in expression (8), an electric current $\mathbf{j} = Ne(\mathbf{v}_i - \mathbf{v}_e)$, perpendicular to magnetic field $\mathbf{B}$ ($\mathbf{j}_\perp$), is straightforwardly derived:

$$(9) \qquad \mathbf{j}_\perp = \sigma_\perp (\mathbf{u}_0 \times \mathbf{B}),$$

where the 'drag' conductivity $\sigma_\perp$ coincides with (7).

Expressions (6-9) enlighten the relationship between the perpendicular conductivity $\sigma_\perp$ and the drag force $\mathbf{F}_{i,e}$ that exerts motion of charged particles in the presence of neutrals fluids with velocity $\mathbf{u}_0$.

## 4. Discussion

Under static conditions the neutral velocity $u_n$ becomes

$$(10) \quad u_n = \frac{\tilde{m}_{ni}\nu_{ni}v_i + \tilde{m}_{ne}\nu_{ne}v_e}{\tilde{m}_{ni}\nu_{ni} + \tilde{m}_{ne}\nu_{ne}}$$

Substituting ion and electron velocities, $v_i$ and $v_e$, neutral velocity (3a) becomes:

$$(11) \quad u_n = \frac{1}{B}\left(\frac{\nu_{in}}{\Omega_i(1+m_i/m_n)} - \frac{\nu_{en}}{\Omega_e(1+m_e/m_n)}\right)(u_0 \times B) \cong \frac{\nu_{in}}{\Omega_i(1+m_i/m_n)}\frac{u_0 \times B}{B}$$

It follows that the ion drag force changes the neutral wind velocity and becomes perpendicular to the initial velocity $u_0$, i.e. the neutral wind will rotate completely to 90° due to the inverse action of the ions on the neutrals. Its magnitude appears proportional to the ratio $\nu_{in}/\Omega_i$, provided that the initial velocity $u_0$ is uniform and constant. This ratio, of course, will depend on height. Its magnitude however is limited from both below and above, because the plasma density distribution in the E region usually is limited in height and lies between 100-125 km. Therefore, the ratio $\nu_{in}/\Omega_i$ and subsequently, the neutral velocity (10) may alter its value in wide range and gets values less and higher than $u_0$.

Concerning the plasma-neutrals system the following conclusions are drawn. At an initial moment when the structure of enhanced plasma density is still not balanced with the neutral wind, the dynamo current system is governed by the well-known Pedersen and Hall electric currents. When the enhanced plasma density structure enters in steady state conditions, the electric currents become aligned perpendicular to neutrals velocity $u_0$ and their magnitude is governed by the perpendicular (drag) conductivity $\sigma_\perp$ and dynamo electric field $u_0 \times B$. Hence, the magnetic field generated by this current (the disturbed magnetic field) would have the same direction as the neutrals velocity $u_0$. Inversely, the neutral wind acquires an additional component $u_n$ which is perpendicular to $u_0$ and thus the neutral wind field becomes re-arranged depending on the collision to ion cyclotron ratio $\nu_{in}/\Omega_i$ magnitude presumably at the plasma density peak.

*Kosch et al* (2010) have present two case studies of high-latitude E region thermospheric winds and their interaction with auroral arcs with meso-scale resolution from Mawson, Antarctica. In these case studies the neutral wind in the E layer is observed to rotate at 90° and to flow parallel to the arc, within a region ~50 km wide on either side of the arc. When the arc disappears the E region winds rotated back, i.e. restores its initial direction (*Kosch et al,* 2010). After having checked whether the observed neutral wind rotation effect is due to changing altitude of the optical emission or to the solar heating pressure and Coriolis forces normally dominant in the E region, *Kosch et al* (2010) concluded that the increased plasma density *and* electric field in the immediate vicinity of the auroral arc and the associated ion drag forces are possible sources of the observed neutral wind rotation. In a support of this assumption the time scale of the ion-neutral coupling is involved. This time scale can easily be derived from Eq. (3). From Eq. (3) the drag force effect on neutrals can be accomplished after a time $\tau_{char}$ (*e*-folding time) equals to:

$$(12) \quad \tau_{char} \approx (N_n/N)m_n/(\tilde{m}_{ni}\nu_{in} + \tilde{m}_{ne}\nu_{en}).$$

Obviously, this result practically coincides with the inverse of the neutral–ion collision frequency $\nu_{ni}$ (the relationship $N_n\nu_{ni} = N\nu_{ni}$, is used). Note *Nozava and Brekke* (1995) and *Kosch et al* (2010) who have taken into account only the ion drag on neutrals, have obtained similar result. Obviously, this frequency is directly governed by the plasma density variations $N$, while the neutral density $N_n$ does not influence $v_{ni}$.

According to *Kosch et al* observations (2010) the observed arc events the peak E region frequency alters from 2.3 to 11.3 MHz (corresponding to $6.5 \times 10^{10}$ and $1.6 \times 10^{12}$ m$^{-3}$) and from 3.1 to 7.1 MHz (corresponding to $1.2 \times 10^{11}$ and $6.2 \times 10^{11}$ m$^{-3}$). Hence, the time scale will decrease by the same amount. Indeed, in the two events the neutral wind vector rotation occurred within one time step, i.e., 7 min, consistent with the large 24 times increase in E region plasma density.

Collision frequency of ions with neutrals has been examined by *Nygren et al.* (1987, 1989) and Davies et al (1997). Conventionally, the ion−neutral collision frequency has been estimated from observations by incoherent scatter radars adopting methods elaborated by *Nygren et al.* (1987, 1989) and *Davies et al* (1997). Based on these methods, the ion−neutral collision frequency has been successfully estimated to altitudes approaching 140 km. The estimated ion−neutral collision frequency varies roughly between $10^2$ Hz to $10^3$ Hz depending on height (109-125 km) (*Davies et al*, 1997). Using these experimental evidences and taking into account the regular stratification changes of the neutral density (in height), the time scale (for E region density conditions of $1.6 \times 10^{12}$ m$^{-3}$) may reach a minimum of 10 minutes and less. Therefore, the estimation of the time scale qualitatively confirms the experimental evidence of 7 min pointed out by *Kosch et al* (2010).

Note that Kosch et al (2010) have argued that their observational fact is consistent with an earlier observation by *Wescott et al.* (2006), who also found (from a rocket chemical release experiment) the E region neutral wind to be parallel to an auroral arc. In fact, a closer examination shows that in the Wescott et all (2006) experiment the neutral wind velocity was directed along the arcs at all stages (before, during and after the arcs event), i.e. no neutral wind rotation effect has been registered there. Presumably the Wescott et all experiment (2006) might be explained by a polarization electric field that subsequently emerges across the arc boundaries.

Another outcome of the analysis should be pointed out. The Pedersen and Hall currents in the E region are workable under a simple condition of immobile neutrals, i.e. for time scales less than $\tau_{char}$ (12). On the contrary, for time scales beyond $\tau_{char}$ the application of the drag conductivity and the associated perpendicular (drift) current is substantiated. The recent observations of neutral wind meso-scale variations in the neutral wind produced by enhanced plasma density arc structures (*Kosch et al.*, 2010) are in favor of this finding.

The plasma−neutral coupling processes that will work effectively in the polar regions of enhanced plasma densities following geomagnetic activity has been highlighted by Lotko (2004). In a support of this expectation a so-called 'flywheel effect' has been pointed out in the high-latitude ionosphere-thermosphere system (Lyons et al., 1985). This 'flywheel effect' is produced through the polarization electric field ($\boldsymbol{E} \rightarrow \boldsymbol{u} \times \boldsymbol{B}$) and perpendicular current carried by collisionally entrained ions.

A further elaboration of the above theoretical finding is that other factors e.g. spatial scales of the ionosphere−thermosphere system, possible emergence of field-aligned electric currents (along the magnetic field $\boldsymbol{B}$) and/or polarization electric fields due to absence of closure of Hall/Pedersen currents along some direction (e.g. in height) must additionally to be taken into account.

**5. Conclusion**

The neutrals in the Earth environment are in fact free and subjected to drag forces (by ions). In particular, the Pedersen and Hall conductivities dos not render an account of such drag forces and are, in principle, applicable under a simple condition of immobile neutrals. Hence, these conductivities are mistakenly applied to electric currents in the Earth's ionosphere and thermosphere and also, to Joule heating evaluations there.

In this study an analysis of the drag forces due to neutral wind, drag conductivity and their relevance to the Earth's ionosphere and thermosphere is performed. A drag conductivity perpendicular to the neutral wind field of velocity $\boldsymbol{u}_0$ and the static magnetic field $\boldsymbol{B}$ is subsequently introduced and derived. The perpendicular electric current expressed through the drag forces is evaluated quantitatively.

The mechanism of this perpendicular electric current is the following. The drag forces in the ionosphere−thermosphere system initiate changes in the plasma flow, neutral wind. In the presence of initial neutral wind field of velocity $\boldsymbol{u}_0$ the ions and electrons acquire drifts perpendicular both to the initial neutral wind velocity and to the ambient magnetic field producing subsequently a perpendicular electric current. This perpendicular electric current is defined by a drag conductivity and the polarization electric field $\boldsymbol{u}_0 \times \boldsymbol{B}$. Self-consistently, the free neutrals acquire an additional neutral velocity component perpendicular to the initial neutral wind velocity $\boldsymbol{u}_0$. This process inevitably is accompanied by fading of the Pedersen and Hall currents that wane within a specific time. This time scale is inversely proportional to neutral-ion collision frequency.

The theoretical results supposedly are supported by a recent observation of meso-scale neutral wind rotation in the vicinity of auroral arcs (Kosch et al, 2010).

These findings are relevant to a better understanding of electric current generation, distribution and closure in weakly ionized plasmas where charged particles (plasma) and neutrals are not bound (free) as the ionosphere and thermosphere in the Earth environment, the solar atmosphere-chromosphere region and even dusty plasmas.